# Deciphering the dynamics of Epithelial-Mesenchymal Transition and Cancer Stem Cells in tumor progression

Federico Bocci[1,2], Herbert Levine[1,2,3,4], José N Onuchic[1,2,4,5], Mohit Kumar Jolly[1,6,#]

[1]Center for Theoretical Biological Physics, Department of [2]Chemistry, [3]Bioengineering, [4]Physics and Astronomy, [5]Biosciences, Rice University, Houston, TX, USA 77005

[#]Corresponding author:
[6]Current address: Centre for BioSystems Science and Engineering, Indian Institute of Science, Bangalore, India 560012. Email: mkjolly@iisc.ac.in

## Abstract

**Purpose of review:**
The Epithelial-Mesenchymal Transition (EMT) and the generation of Cancer Stem Cells (CSCs) are two fundamental aspects contributing to tumor growth, acquisition of resistance to therapy, formation of metastases, and tumor relapse. Recent experimental data identifying the circuits regulating EMT and CSCs has driven the development of computational models capturing the dynamics of these circuits, and consequently various aspects of tumor progression.
**Recent findings**:
We review the contribution made by these models in a) recapitulating experimentally observed behavior, b) making experimentally testable predictions, and c) driving emerging notions in the field, including the emphasis on the aggressive potential of hybrid epithelial/mesenchymal (E/M) phenotype(s). We discuss dynamical and statistical models at intracellular and population level relating to dynamics of EMT and CSCs, and those focusing on interconnections between these two processes.
**Summary:**
These models highlight the insights gained via mathematical modeling approaches, and emphasizes that the connections between hybrid E/M phenotype(s) and stemness can be explained by analyzing underlying regulatory circuits. Such experimentally curated models have the potential of serving as platforms for better therapeutic design strategies.

# Introduction

Metastasis and tumor relapse are insuperable clinical challenges that claim most cancer-related deaths [1]. The metastatic cascade has extremely high rates of attrition, because of the multi-step and challenging sequence of events leading to a secondary tumor. These steps include the detachment of cancer cells from their home organ, their circulation in the bloodstream, and eventually their colonization of a foreign environment, all while escaping attack by the immune system and other clinical interventions.

A first step in the metastatic cascade is a phenotypic switch called Epithelial-to-Mesenchymal Transition (EMT). Cancer cells in a solid tumor tissue often undergo EMT, characterized by the loss of cell-cell adhesion and acquisition of migratory and invasive traits [2]. Disseminated cells travel through the bloodstream and colonize a distant organ, giving rise to macrometastases [2, 3]. EMT is not necessarily a cell-autonomous and binary process. Cells can attain one or more intermediate, or hybrid, epithelial/mesenchymal (E/M) state(s) and can involve their neighbors to form more aggressive clusters of circulating tumor cells (CTCs) - the main drivers of metastases [4–6]. EMT is regulated at multiple levels – transcriptional, translational, post-translational and epigenetic – by many context-specific factors and the tumor microenvironment. Some common traits of EMT include transcriptional repression of E-Cadherin that mediates cell-cell junctions and adhesion, and activation of one or more EMT-inducing transcription factors (EMT-TFs) such as SNAI1, SNAI2, ZEB1, ZEB1, TWIST1 that induce can cell scattering, motility, and invasion [2].

To colonize a secondary tumor site, the disseminated tumor cells need to give rise to different cell types that constitute a tumor – a trait typical of cancer stem cells (CSCs). Cells with such stem-like properties are also typically resistant to various clinical treatments, and are often implicated in tumor relapse. The conventional, so-called 'CSC hypothesis' envisions a small fraction of CSCs that can both self-renew (symmetric division) and generate differentiated cells (asymmetric division) [7, 8]. This hypothesis implies a hierarchical lineage of tumor cells similar to stem cell hierarchy in normal tissues, such that CSCs that differentiate irreversibly lose their stem-like properties [9]. Recent studies, however, have emphasized that stemness can be a dynamic cell state that can be acquired or lost [10–13]. In other words, some differentiated tumor cells can dedifferentiate and regain stemness via epigenetic and/or environmental factors such as abnormal cancer metabolism and EMT.

The interconnection between EMT and CSCs was first postulated by Brabletz *et al.* [14] in 2005 as 'migrating cancer stem cell' by suggesting that the concepts of EMT and CSCs, considered independent of one another, were not sufficient to explain various traits of cancer progression. Afterwards, experimental evidence accumulated suggesting that stemness can be gained during EMT [2, 5, 15–19]. Recent experiments have shown that cells in intermediate E/M states possess a higher metastatic potential as compared to the cells that have undergone a complete EMT. Moreover, cells in hybrid E/M phenotype have been suggested to be drug-resistant [20, 21]. Put together, these observations emphasize the clinical implications of hybrid E/M phenotypes [5, 22, 23].

Recent studies have made significant progress in identifying the molecular networks regulating EMT, CSCs, and their interconnections [24]. These networks are formidably complex, and capable to give rise to emergent non-linear behavior. Identification of these networks has driven a surge in deciphering their underlying principles from a dynamical systems perspective. This approach has involved developing many computational models to capture the dynamics of these transitions. These models may deal with intracellular and intercellular circuits, or may offer a population level description without considering the detailed dynamics of signaling networks. Here, we review both of these types of models. First, we review a set of models that attempt to characterize the possible set of states for cells undergoing EMT and their possible relevance to tumor progression and metastasis. Second, we review a set of models that consider the population structure of a tumor and its implications for drug resistance. Finally, we discuss models that aim at gaining a comprehensive understanding of the connection between these two crucial axes of cancer progression.

## Mathematical models of EMT

Computational models developed for EMT can be categorized broadly into two classes: mechanism-based models and data-based models. While the first class of models adopts a 'bottom-up' approach and focus on elucidating the properties of molecular networks identified experimentally, the latter adopts a 'top-down' approach starting with high-dimensional data and aims to reverse engineer the networks, and/or trace the trajectories of these transitions using statistical methods.

1. Decoding the dynamics of cellular transitions: Mechanism-based models of EMT

The first set of mechanism-based models for EMT regulation – developed independently by two groups – focused on a small set of nodes, and captured the dynamical features emerging from the interconnections among those nodes (Fig. 1A, left). These models included the EMT-suppressing microRNA families miR-34, miR-200 and the families of EMT-TFs ZEB and SNAIL [25, 26]. Both models predicted that this network can be tristable, and could give rise to a hybrid epithelial/ mesenchymal (E/M) phenotype, in addition to epithelial and mesenchymal phenotypes (Fig. 1A, right) [25, 26]. These models also suggested that more than one phenotype can be accessible to a cell due to the underlying multistability, hence giving rise to sub-populations of epithelial, hybrid E/M and mesenchymal cells in a genetically identical population. This phenomenon was observed and later characterized in detail in multiple cancer cell lines [5, 27–29]. Due to different modeling approaches, however, these models differed on the dynamics of attaining this hybrid E/M phenotype. Experimental support for both these models has been observed [27, 30], highlighting the heterogeneity and multiplicity of hybrid E/M phenotype(s) present in different cell lines.

Further follow-up work has identified several intracellular phenotypic stability factors (PSFs) that can stabilize a hybrid E/M phenotype, including OVOL2, GRHL2, Np63α and NRF2 [31–35]. Their role as PSFs have also been validated experimentally *in vitro* and *in vivo* [32, 34–36]. Moreover, higher levels of these PSFs were observed to correlate with worse patient survival, emphasizing the clinical implications of hybrid E/M phenotype(s) [4, 22]. Among those, NRF2 has been specifically proposed to be maximally expressed in hybrid E/M phenotype(s) [35]. In addition, different energy landscape approaches have been also developed for the aforementioned EMT circuit [37] as well as for related larger gene regulatory circuits [38]. This strategy allows to compute the transition rates between multiple cell states, and thus predict the relative abundance of different phenotypes (epithelial, mesenchymal, and hybrid E/M) in an isogenic population.

EMT can be also induced by biochemical signals coming from neighboring cells. Boareto *et al.* [39] elucidated the connection between EMT and the Notch signaling pathway, a cell-cell, contact-based, evolutionary conserved signaling mechanism that is also implicated in angiogenesis and therapy resistance. The model predicted that Notch-Jagged signaling among cells, but not Notch-Delta signaling, can foster the formation of clusters of hybrid E/M cells by promoting a similar hybrid E/M phenotype in neighboring cells [39]. Consistently, gene expression analysis highlighted higher levels of Jagged in CTC clusters of patients as compared to single CTCs [33]. Thus, a hybrid E/M phenotype can be stabilized not only by intracellular PSFs directly coupled to the EMT core circuit, but also via cell-cell signaling. As another example, Bocci *et al.* [40] predicted that Numb – an inhibitor of Notch signaling – can also stabilize a hybrid E/M phenotype; this prediction was validated experimentally in multiple independent studies [40, 41].

As the network grows in size (such as going from Fig 1A to Fig. 1B), identifying kinetic parameters becomes more and more challenging. Computational models that have focused on such larger networks have typically been simulated using Boolean modeling approaches, where the state of gene expression is either On (active) or Off (inactive). Boolean models do not consider any kinetic parameters. Cohen *et al.* [42] developed a Boolean network to evaluate the combinatorial effect of different mutations on EMT and metastatic potential using transcriptome data from TGF-$\beta$-induced EMT. Similarly, Steinway *et al.* [43] constructed a circuit for TGF-$\beta$-induced EMT using data from hepatocellular carcinoma (HCC). Their model predicted the activation of several pathways during EMT such as Sonic Hedgehog and Wnt. Following up, the authors showed that certain perturbations could give rise to one or more hybrid E/M states, and identified possible targets to inhibit TGF-$\beta$-driven EMT [44]. Recently, Font-Clos *et al.* [45] constructed Boolean model for a gene regulatory network that describes both EMT and its reverse, Mesenchymal-to-Epithelial Transition (MET). An energy landscape approach showed two main attractors, or stable states, corresponding to epithelial and mesenchymal phenotypes, and multiple local minima, or relatively less stable states, corresponding to multiple hybrid E/M phenotypes (Fig. 1C). The authors further mapped RNA-seq data from both lung adenocarcinoma and embryonic differentiation during EMT/MET and compared it to the predicted phenotypic expression profiles, hence validating the existence of multiple different intermediate E/M states.

In an attempt to combine the advantages of both continuous small-scale models and Boolean large-scale models, Huang *et al.* [46] devised an algorithm - Random Circuit Perturbation (RACIPE) - where the expression levels of genes are continuous, but the parameters for all regulatory links are randomly chosen within a biologically-relevant range. RACIPE generates an ensemble of mathematical models, each with a different set of parameters, and identifies the robust dynamical states emerging from a given network topology. Applying RACIPE to an EMT circuit composed of 9 microRNAs and 13 TFs (Fig. 1B) highlighted two different hybrid E/M states [46] that could be stabilized further by stochasticity or noise [47].

2. Reconstructing EMT plasticity from experiments: data-driven approaches to EMT

Recent experimental techniques are capable of generating large and high-throughout ('omics' level) data. This deluge has driven a class of data-driven, or 'top-down', models, which employ a variety of statistical tools to reconstruct correlations among genes and develop expression signatures of different EMT phenotypes.

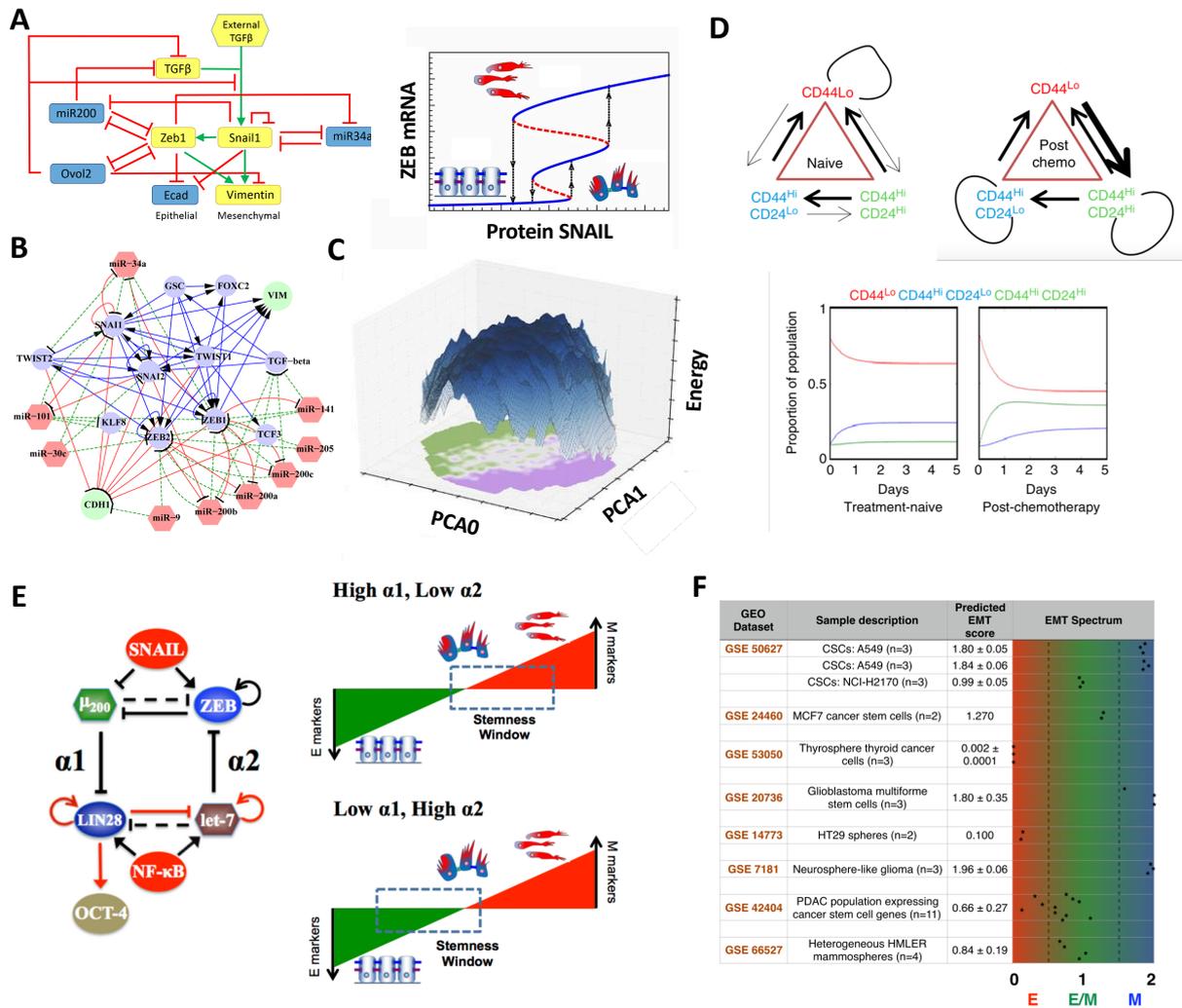

*Figure 1. **Mathematical models that characterize the landscape of cellular plasticity mediated via EMT and CSCs.** **(A)** Left: a gene regulatory circuit for EMT adapted from Hong et al. [34]. Right: a bifurcation diagram of ZEB mRNA as a function of EMT-TF SNAIL adapted from Lu et al. [25] shows three stable phenotypes (i.e. continuous black curves) corresponding to epithelial (low ZEB), hybrid E/M (intermediate ZEB) and mesenchymal (high ZEB). **(B)** An extended EMT regulatory circuit adapted from Huang et al. [46]. **(C)** The energy landscape of a large EMT regulatory circuit adapted from Font-Clos et al. [45] shows two main minima (purple and green projections) corresponding to epithelial and mesenchymal phenotypes, respectively. Additionally, many local energy minima en route to EMT correspond to intermediate E/M states. PCA0 and PCA1 are the first two components of the Principal Component Analysis of the circuit. **(D)** Chemotherapy increases the population of chemo-resistant cancer cells (CD44$^{hi}$CD24$^{hi}$) by increasing the conversion rate from low-resistance CD44$^{low}$ cell population. Top: circuit schematic; Bottom: Temporal dynamics of cancer cell subpopulations pre- and post-treatment adapted from Goldman et al. [67]. **(E)** Left: a core gene regulatory circuit including regulation of EMT via the miR-200/ZEB axis and stemness via the LIN28/let-7 axis adapted from Jolly et al. [12]. The parameters $\alpha_1$, $\alpha_2$ represent the strength in the regulation of the stemness circuit by the EMT circuit ($\alpha_1$) and the EMT circuit by the stemness circuit ($\alpha_2$). Right: varying $\alpha_1$ and $\alpha_2$ shifts the 'stemness window' along the EMT axis adapted from Jolly et al. [12]. **(F)** The EMT score of different cancer stem cell lines adapted from Bocci et al. [13] shows the spread of CSC properties along the EMT spectrum. The score classifies cells as epithelial (score<0.5), hybrid E/M (0.5<score<1.5) or mesenchymal (score>1.5). Each row depicts a different CSC line, and each dot depicts a different biological replicate.*

For example, Zadran *et al.* [48] analyzed the temporal mRNA data of A549 lung cancer cells treated with TGFβ, and identified an intermediate EMT state with a metabolic state characterized by increased cytosolic ATP levels [48]. Further, Chang *et al.* [49] analyzed the time course data of TGF-$\beta$-driven EMT for A549 cells and identified three master TFs for a partial EMT state - ETS2, HNF4A and JUNB [49]. These regulators correlate with a worse clinical outcome and their knockdown can prevent TGFβ-driven EMT [49], reminiscent of observations made for PSFs.

Besides, two different groups developed methods to analyze gene expression data of a certain cell line or tumor cell and calculate an 'EMT score' to quantify the positioning of these cells along the EMT spectrum. The algorithm developed by Tan *et al.* uses entire transcriptomic data for a given sample [50], while that developed by George *et al.* considers a few E and M markers (such as E-Cadherin, Vimentin etc.) as well as PSFs of hybrid E/M phenotypes (OVOL, GRHL2 etc.) [23].

Data-driven models do not necessarily rely on omics-level data; they can also use morphological data. For instance, Mandal *et al.* [51] proposed a phenomenological approach to elucidate intermediate EMT states based on cell microscopy during EMT, and found 3 intermediate states with different morphological attributes [51]. A more rigorous analysis was proposed by Leggett *et al.* [52] that relies on single cell microscopy to classify cells as epithelial or mesenchymal with high precision during TGF-β-driven EMT [52]. Even further, Zhang et al. [53] classified the morphology and motility of migrating breast cancer cells using machine learning algorithms such as Artificial Neural Networks (ANN) and Random Decision Forest (RDF) to analyze single-cell microfluidic microscopy images [53].

As the connections between molecular and morphological traits of EMT continue to be explored in detail [54], a synergistic crosstalk among the computational models described above and their integration with experimental data can provide novel and crucial insights into the dynamics of EMT.

## Mathematical models of CSCs

1. **Modeling the CSC fraction during tumor progression**

An important direction where mathematical approaches have offered significant insights into the CSC dynamics and its relationship with tumor progression is a set of population dynamics models that aim at understanding the temporal dynamics and the mechanisms regulating the CSC fraction (the fraction of cells with stem cell properties in a tumor) [55, 56]. Dhawan *et al.* [57] considered two compartments, the CSC-like cells and the non CSC-like cells, to elucidate the increased plasticity observed in human mammary epithelial cells under hypoxia. In their model, individual cells can both differentiate (from CSC to non-CSC) and dedifferentiate (from non-CSC to CSC). Integrating their model with gene expression analysis, the authors showed that hypoxia generates a shift toward a more CSC-like population and increases EMT features [57].

The role of cell dedifferentiation to a stem-like state has been also investigated by Jilkine *et al.* [58] via a hybrid model that describes the development of the differentiated cell population in a deterministic manner, but considers the stochastic accumulation of mutations to better describe the small CSC population. The authors concluded that dedifferentiation to a stem-like state can speed up tumor progression by enlarging the CSC population [58].

Among the different possible mechanism for dedifferentiation, metabolic reprogramming is especially frequent in the context of cancer [59]. Liu *et al.* [60] devised a probabilistic framework to specifically investigate metabolic reprogramming that converts somatic cells into pluripotent stem cells. Insights gained from such analysis may be useful in understanding metabolic aspects of tumor cell dedifferentiation, given that different subsets of CSCs may have different metabolic vulnerabilities [61].

Another set of models have aimed to explain how a fraction of CSCs is maintained in a tumor. Zhou *et al.* [62] developed a population model of tumor growth that integrates the differential growth rate of CSCs and differentiated cells, as well as transitions among cell phenotypes. In particular, switching between phenotypes maintains a fixed ratio of cell sub-populations [62]. Extending this idea, Wang *et al.* [63] proposed a population model that combines hierarchical organization (irreversible loss of stem traits upon differentiation) and stochastic switching (stemness can be gained by switching to a stem-like state). In this model, CSCs can (a) self-renew (symmetric division in two CSCs), (b) differentiate (symmetric division into two non-stem cells) and (c) asymmetrically divide into a CSC and a daughter differentiated cell. Additionally, differentiated cells can proliferate (symmetric division) but also switch to the progenitor CSC state [63]. The combination of hierarchical and stochastic processes can reproduce the CSC/differentiated cell fraction observed in a human colon cancer cell population [63]. A similar idea has been proposed by Zhou *et al.* [64] to show that back and forth transitions between stem-like and non-stem states is crucial to establish an equilibrium cell fraction of CSCs [64].

A different approach to model CSC-driven tumor progression was proposed by Poleszczuk *et al.* [65]. They proposed an agent-based model where CSCs can gain migratory traits by stochastic mutations. Such approach enables to simulate the spatiotemporal dynamics of the cancer cell population and investigates the cell heterogeneity that arises during tumor development due to mutations. In this model, CSC can divide symmetrically or asymmetrically, and also have a migration potential that translates into discrete movements on a two-dimensional lattice [65], reminiscent of the idea of 'migrating cancer stem cell' proposed by Brabletz [14].

2.   **CSC, tumor progression and therapy: from modeling to the clinic**

A subset of recent models have focused their attention toward identifying optimal therapy schedules for cancer treatment [66]. In this context, CSCs are considered as important target because they are resistant to therapy, and can therefore drive tumor progression and relapse. For instance, the model developed by Dhawan *et al.* [57] (discussed in the previous section) can be generalized to the context of drug tolerance by introducing one or more additional cell sub-populations that can resist different treatments [67]. Specifically, the authors show by integrating *in vivo* experiments and mathematical modeling that chemotherapy can change the rates of conversions among different cell phenotypes and promote a chemotherapy-tolerant state (Fig. 1D) [67].

A more data-driven approach aims to correlate the CSC population with tumor progression and response to therapy. For instance, Werner *et al.* [68] proposed a computational method to quantify the fraction of tumor-initiating cells (i.e. CSCs) by analyzing the tumor's macroscopic growth rate as a function of time. This patient-specific method can be applied to many types of tumors, and provides an estimate of the CSC fraction to rationalize the optimal therapy in a clinical setting [68]. Zhou *et al.* [69] applied a statistical approach to compute the transition rate between CSC and differentiated cells in colon cancer cells and showed phenotypic plasticity with back and forth transitions [69]. Furthermore, Yu *et al.* [70] gathered the differential response of CSCs and differentiated cells to radiotherapy for different tumor types including glioblastoma, lung, prostate, and breast cancer, and fitted this tumor-specific information with a stochastic mathematical model to explain the different inter-tumor responses to radiation therapy [70].

Not all models of cancer cell-therapy interplay need to employ a population approach. Instead, Chen *et al.* [71] used an energy landscape approach to investigate the transitions of breast cancer cells which are sensitive, hypersensitive or insensitive to hormone therapy regulated by the ER$\alpha$ signaling network. The authors implemented different treatment strategies including sequential treatment (multiple drugs) and intermittent treatment (alternation of treatment and 'drug holiday' periods) [71]. The effects of continuous vs. intermittent treatments was also explored in the context of prostate cancer, where a small-scale model predicted that cells could oscillate between a therapy-sensitive and a therapy-resistant phenotype [72]. The authors further modeled different hormonal treatments for prostate cancer that were predicted to synchronize oscillations among different cells, thus restricting the heterogeneity in the cell population [73].

As discussed above, most models related to CSCs have largely focused on identifying the causes underlying varying fractions of tumor population that can behave as CSCs [74]. Thus, there is much room for progress in constructing mechanistic models for CSC-driven tumor progression and the emergence of drug-resistant phenotypes. In this direction, Nazari *et al.* [75] recently proposed a mathematical model for the role of inflammatory cytokines in mediating CSC-driven tumor growth. This model couples the ligand-receptor interaction at the molecular scale with CSC self-renewal and proliferation at the cellular level [75], and could reproduce the observed decrease in tumor volume in mouse models with knockdown of IL-6R, an inflammatory cytokine.

## Towards an integrated understanding of EMT and CSC

Aside from the separate models for EMT and CSC dynamics as discussed above, multiple computational models have investigated the connection between EMT and CSC. Turner *et al.* [76] interrogated the connection between EMT and CSC through a phenomenological, population model with two possible scenarios where EMT enriches the CSC population. First, cells can dedifferentiate back to a CSC state while undergoing EMT. Secondly, EMT increases the probability of symmetric, self-renewal division of cells that are already stem-like [76]. The authors used the model to fit the experimental data on CSC fraction and mammosphere expansion, indicating that both processes may play an important role in supporting cancer progression [76]. Later, Gupta *et al.* [77] showed that breast cancer cells can exist in different sub-populations with varying functional attributes: luminal, basal and stem-like. They demonstrated that the overall population, when perturbed, re-establishes a fixed fraction of the three cell phenotypes. This robustness could be explained by a population model where cells can undergo stochastic phenotypic transitions between the three different states [77]. Moreover, the stem-like cell line SUM 149 has been shown to exhibit the traits of a hybrid E/M phenotype, hence suggesting a possible correlation between a partial EMT state and stemness [33].

The multi-scale model proposed by Sfakianakis et al. [78], instead, focuses on resolving the spatial structure of a cancer cell population. This phenomenological model couples the aspects of CSC and EMT to describe the invasion of extracellular matrix by tumor cells. In this framework, EMT is modeled as a binary switch between an epithelial-like and a mesenchymal-like phenotype that is driven by growth factors. Therefore, this model couples EMT at the individual cell scale and the population dynamics and growth of the tumor mass at the multi-cell scale. Note, however, that the models discussed so far proposed mechanisms for CSC-driven tumor progression and maintenance of the CSC fraction, but did not provide a molecular rationale for the acquisition of CSC traits.

Li and Wang [38] reconstructed a core gene regulatory circuit with relevant players determining CSC properties such as miR-145 and OCT4, and core regulators of EMT - miR-200 and ZEB. The authors applied an energy landscape approach to predict the co-existence of multiple cellular phenotypes. In their model, a cell can either assume a 'normal' state or a 'cancer' state, both of which could or could not exhibit stem-like traits. Thus, a total of four possible cell phenotypes are available – normal, normal stem-like, cancer, and cancer stem-like [38]. In this framework, p53 represents a degree of cancerization and ZEB represents a degree of stemness. Notably, both the predicted 'normal stem cell' state and the 'CSC' state highly express ZEB, hence implicitly suggesting that stemness is gained along with EMT [38].

Finally, the models developed by Jolly and colleagues explicitly proposed a mechanism-based rationale to elucidate the connection between EMT and CSC: the stemness circuit comprising LIN28, let-7, and OCT4 is connected to the EMT circuit already discussed by Lu et al. [25] (Fig. 1E, left). The CSC phenotype was defined as a state with intermediate levels of OCT4 that have been shown experimentally to correlate with stem-like traits [79, 80]. These models proposed that a CSC phenotype is highly correlated with a hybrid E/M phenotype [81], but intracellular factors such as OVOL [12] or cell-cell communication via Notch signaling [13] could move the predicted 'stemness window' toward the epithelial or mesenchymal ends of the EMT spectrum (Fig. 1E, right). Experimental evidence for this dynamic 'stemness window' concept was provided by Bocci et al. [13] by computing the 'EMT score' [23] of different human CSC lines using publicly available datasets. This analysis showed that CSC traits can be scattered along the EMT spectrum based on context-specific activation of signaling pathways, therefore resulting in epithelial, hybrid E/M and mesenchymal CSC (Fig. 1F) [13]. Furthermore, this model proposed a strong overlap between a hybrid E/M phenotype, CSC properties and Notch-Jagged signaling [13], a pathway implicated in both drug resistance and in clusters of CTCs, the key drivers of metastasis [33]. Consistently, knockdown of Jagged was shown to restrict the growth of tumor emboli in SUM149 inflammatory breast cancer cells [82]. Given the role of Notch signaling in pattern formation in multiple contexts [83, 84], Notch signaling coupled with EMT circuitry may underlie the spatial segregation of different subsets of cells with stem-like traits, as observed experimentally in a breast cancer tissue [85]. Secretion of a diffusive EMT-inducing signal at the tumor-stroma interface (such as TGF-$\beta$), along with cell-cell signaling through Notch, was shown to give rise to mesenchymal CSCs at the invasive edge of the tumor and a population of hybrid epithelial/mesenchymal (E/M) CSCs in the tumor interior [82]. The idea that cell-cell signaling and the microenvironment can shape the spatial distribution of a cell population has been examined in different biological contexts, such as bacterial colonies [86, 87] or eukaryotic chemotaxis [88, 89], but remains largely unexplored in cancer biology, and thus demands further attention.

## Conclusion

EMT and CSC represent two crucial biological axes that bolster tumor progression, metastasis and tumor relapse [2, 4]. While the molecular details of multiple steps of tumor development continue to be identified, it is largely accepted that EMT often plays a crucial role in regulating epigenetic, morphological and functional cell properties during tumor progression and metastasis formation [2, 4]. Similarly, it is well accepted that the acquisition of stem-like properties potentiates tumor maturation and enhances resistance to various treatments, driving tumor relapse. Only recently, we have been gaining insights into how, when and where these two dynamic processes can influence one another (Fig. 2). In this context, mathematical modeling has proven itself as a potent tool to interpret existing data and formulate new predictions that can be tested experimentally.

In the context of EMT, mechanism-based computational models have suggested that cells undergoing EMT can stably acquire intermediate cell states enabling hybrid phenotypes with mixed epithelial (E) and mesenchymal (M) characters, as opposed to a binary E-M switch scenario [2, 4]. Novel *in vivo* and *in vitro* analysis recently highlighted the existence of such hybrid states that coexpress E and M markers and often possess mixed morphological traits of cell-cell adhesion and motility [18, 19], and have highlighted their enhanced metastatic potential [90]. The next crucial steps will include a more comprehensive attempt to integrate data-based models, mechanism-based models, and time-course and single-cell experimental data, to formulate a more quantitative characterization of these malignant hybrid E/M state(s).

In the context of CSC, one set of models considers the dynamics of a CSC population employing the tools of population dynamics and agent-based modeling. Such class of models can provide predictions about CSC fraction or population dynamics under perturbations, hence potentially providing strategies for containing CSC-driven tumor progression. Additionally, coupling mathematical modeling with clinical data of therapy response enables predictive tools that can shed light on the CSC-therapy interplay. Such models can provide information on, among others, adaptive response, differential drug sensitivity, or phenotypic plasticity in a cancer cell population.

Recent experimental observations have led to a class of models that can offer insights into the coupling between EMT in cancer cells and the acquisition of stem-like properties. A first set of models relates phenomenologically the acquisition of stem traits with the EMT process, hence explaining how CSC-EMT interaction can support tumor progression and maintain a certain fraction of different cell phenotypes. Moreover, a second class of models investigates the coupling between EMT and CSC at the level of gene regulatory networks, showing a correlation between the cell phenotypes enabled by an EMT regulatory circuit and the stemness regulatory circuit. A common feature across these models is envisioning the acquisition of stemness as a dynamical process correlated with EMT [12, 13, 38]. Recent mathematical modeling and experiments have suggested a correlation among hybrid E/M states and stem cell properties [5, 12, 13, 18, 19]. CSC traits, however, are not exclusively observed in intermediate states, and the crosstalk between tumor, micro-environment and therapies is likely to play a major role in modulating the plasticity properties of cancer cells (Fig. 2), as shown by recent experiments highlighting subsets of CSCs in multiple cancer types [20, 85].

Considered together, these computational models developed for EMT, CSC, or their inter-connections have contributed not only in deciphering the mechanisms underlying specific experimental observations, but also have driven the next set of experiments by generating testable predictions. Such bidirectional crosstalk between theory and experiments can significantly accelerate our goal of understanding and consequently targeting these processes for therapeutic benefit.

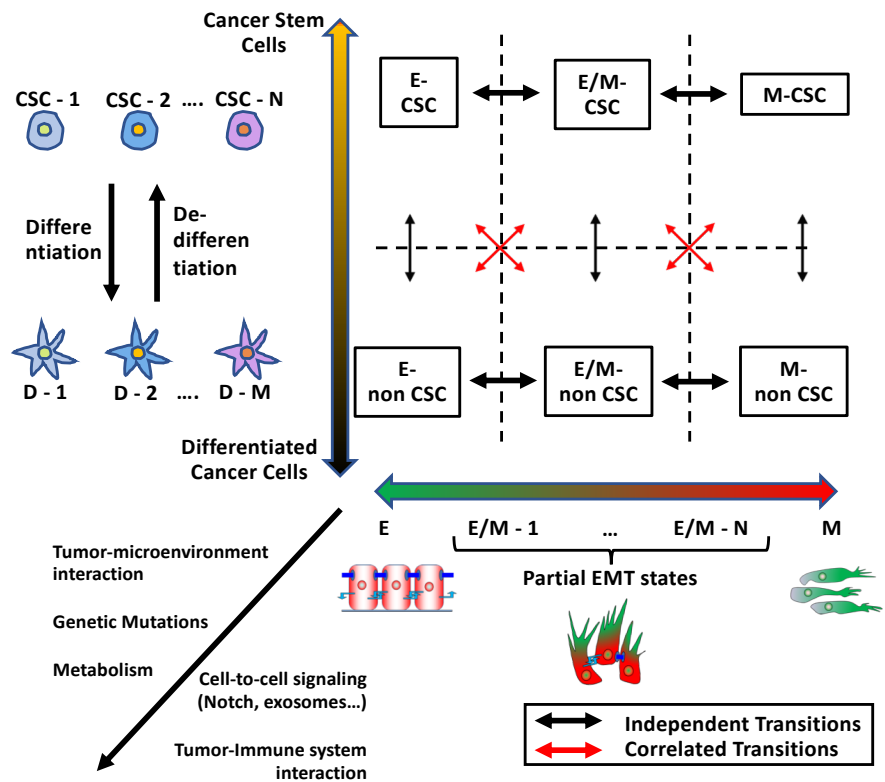

*Figure 2.* **Interconnections between EMT and CSC axes.** *The spectrum of EMT states can range from epithelial (E) to mesenchymal (M) phenotypes, with a variable number of partial E/M states (x-axis). Both cancer stem cells and differentiated cancer cells can assume different states depending on genetic and epigenetic factors (y-axis). Horizontal and vertical transitions represent independent EMT and stem processes, respectively. However, transitions are possible where both the EMT and stem states change, or correlated transition. The interconnection among EMT and CSC states and the transitions enabled between them depend on context-specific factors including, among others, intra-cellular signaling and mutations, cell-cell and cell-environment signaling.*

**Grant support**
This work was sponsored by the National Science Foundation NSF grants PHY-1427654 (Center for Theoretical Biological Physics), PHY-1605817, CHE- 1614101, MCB-1241332 and by the Cancer Prevention and Research Institute of Texas (CPRIT-grant R1110). FB is also supported by the Hasselman fellowship for excellence in Chemistry. MKJ is also supported by a training fellowship from the Gulf Coast Consortia, on the Computational Cancer Biology Training Program (CPRIT Grant No. RP170593).

The authors declare no potential conflicts of interest.